\documentclass[twocolumn]{NobArticle}
\usepackage{color}

\runninghead{Complex ``totopapa'': predicting the successor to pope Francis}

\title{Complex {\itshape totopapa}: predicting the successor to pope Francis}

\author{
    Alberto Antonioni\textsuperscript{1}, 
    Michele Re Fiorentin\textsuperscript{2} 
    and Eugenio Valdano\textsuperscript{3}
}

\date{
    \textsuperscript{\textbf{1}} GISC, Department of Mathematics, Carlos III University of Madrid, 28911 Leganés, Spain\\ 
    \textsuperscript{\textbf{2}}
    Department of Applied Science and Technology, Politecnico di Torino, corso Duca degli Abruzzi 24, 10129 Torino, Italy \\ 
    \textsuperscript{\textbf{3}}
    Sorbonne Universit\'{e}, INSERM, Institut Pierre Louis d'Epid\'{e}miologie et de Sant\'{e} Publique, F75012, Paris, France.\\
    \smallskip
    Contacts: alberto.antonioni@uc3m.es, michele.refiorentin@polito.it, eugenio.valdano@inserm.fr\\
    \medskip
    \today
}

\renewcommand{\maketitlehookd}{%
\begin{abstract}
    \noindent Following the death of Pope Francis, the College of Cardinals will convene in conclave to elect the new Pontiff of the Roman Catholic Church.
    In this report, we present a computational framework for analyzing the ideological landscape of the cardinal electors and estimating the likely outcomes of the upcoming papal election. We collected public textual data describing each cardinal's positions on a range of topics currently relevant to the Church, including sexuality, migration, poverty, governance and interreligious dialogue. These texts were embedded using a transformer-based language model to construct a semantic similarity matrix among cardinals. We then simulated conclave voting dynamics using this matrix as a proxy for ideological proximity. The model produced both topic-conditioned and aggregate probabilistic forecasts of election outcomes. We report the predicted leading candidates under various scenarios and discuss the structure of factional clustering revealed by the embeddings.
    The results highlight a polarized field with a small number of structurally central candidates, among whom Pietro Parolin, the former secretary of state, consistently emerges as the most broadly electable across thematic scenarios, followed by both high-profile and outsider names such as card. Brislin and Tagle.\smallskip\\
    \textbf{Post-election analysis added at the end of the document.}
    \medskip

    \small{\textbf{Index Terms:} Roman Catholic Church, pope, conclave, text embedding, complex systems}
\end{abstract}
}

\begin{document}

\small
\maketitle

\section{Introduction}
On April 21, 2025, Francis (latin: Franciscus), born Jorge Mario Bergoglio and the 266th pope of the Roman Catholic Church, passed away. In accordance with canonical tradition, the College of Cardinals will convene in conclave on May 7 to elect the next Supreme Pontiff. The conclave, held under conditions of strict secrecy in the Sistine Chapel, assembles all cardinal electors --~those under the age of eighty~-- who vote successively until one candidate secures a two-thirds majority.

The choice of a new pope carries consequences beyond the Church. For Catholics, the pope is the supreme authority on doctrine, discipline and governance. For the rest of the world, the papacy is the head of a global institution with moral, cultural, and diplomatic influence. The next pope will shape the Church's response to global issues such as war, migration and climate change, its attitude to ethical and social stances, its internal reform and its relationships to various societies and religions.

As such, the upcoming election is a major geopolitical event and is generating massive speculation on who the next pope will be.
In this report, we join this speculation and construct a data-driven model to estimate the likely outcomes of the upcoming conclave. For each cardinal elector, we collected publicly available textual data describing their positions on a range of topics salient to the contemporary Church. We embedded these texts using a sentence-level transformer model (SBERT), and used the resulting representations to compute a multilayer cardinal--to--cardinal ideological similarity matrix. This ideological map was then used to simulate the dynamics of papal voting under different assumptions.

Our model produced conditional forecasts: given a specific topic as the dominant axis of discourse in the conclave, it predicts which candidate would be most likely to gather the support needed to reach the required two thirds of the votes. Then, we also produced an aggregate forecast by weighing all the topics equally.

This document presents the structure of the method, the clustering of ideological positions, and the resulting probabilistic rankings.

\section{Methods}

\subsection{Data}
Data on the 135 cardinal electors eligible to vote in the upcoming conclave were sourced from  \textit{The College of Cardinals Report} website \cite{coc}, a publicly available and regularly updated web-based resource that compiles biographical and thematic profiles of members of the College of Cardinals. For each cardinal, the available textual content, including official statements, public interventions, interviews, and media summaries, was processed using the large language model ChatGPT \cite{chatgpt}. The purpose of this step is twofold: to generate concise summaries of each cardinal’s public stances and to ensure terminological and thematic uniformity across profiles. We focused on four key areas of interest that reflect prominent debates within the contemporary Catholic Church. These are: (1) the cardinal’s position towards same-sex couples, LGBT persons and related pastoral questions \textit{(Attitude towards LGBT)}; (2) views on synodality, understood as the role of collegial decision-making and consultation within ecclesial governance \textit{(Synodality)}; (3) positions on migration, economic inequality, and poverty \textit{(Migrations and poverty)}; and (4) engagement in the dialogue and outreach with religions beyond Catholicism \textit{(Interreligious dialogue)}. Summaries were generated using consistent prompts designed to extract each cardinal’s position in a comparable and thematically aligned format, while minimizing interpretive bias.

\subsection{Cross-encoding and similarity}
To investigate similarities between cardinals with respect to their thematic positions, we employed the \texttt{stsb-roberta-base} cross-encoder model \cite{sentence-bert}, a pretrained transformer-based architecture fine-tuned on semantic textual similarity (STS) benchmarks. For each of the four selected topics (Attitude towards LGBT, Synodality, Migrations and poverty, and Interreligious dialogue) we applied the cross-encoder to compute pairwise similarity scores between all possible pairs of cardinals in the Conclave. Specifically, for a given topic, we input the corresponding paragraphs of each pair of cardinals into the cross-encoder. The model returns a real-valued similarity score, with values closer to zero indicating greater semantic divergence, and higher values denoting stronger similarity in expressed positions. This procedure yielded a topic-specific similarity matrix, where each entry quantifies the degree of alignment between two cardinals on a particular theme.

On each of these similarity matrices, we applied spectral clustering to identify coherent groups of cardinals with similar thematic profiles. Each resulting cluster was then characterized by aggregating the ideological scores (see below) of its members, allowing us to assess whether the cluster as a whole tends to express more progressive, conservative, or intermediate positions on the corresponding theme.

In addition to computing pairwise similarities among cardinals, we constructed two short synthetic reference texts for each theme, representing a prototypical progressive and conservative position, respectively. We then cross-encoded the thematic paragraph of each cardinal with both reference texts and computed the difference between the resulting similarity scores. This yielded a single scalar value for each cardinal and topic: positive values indicate greater alignment with the progressive stance, while negative values reflect closer similarity to the conservative one. In this way, each cardinal is assigned a position along a continuous axis of ideological orientation for each of the four themes.

For additional comparison, we performed the same cross-encoding procedure using paragraphs specifically prepared to reflect the discourse and pastoral tone of Pope Francis. This allowed us to assess the extent to which each cardinal aligns with the positions that characterized his pontificate.

\subsection{Voting model}

The voting model works in rounds just as the actual conclave. In the first round of voting each cardinal votes for a candidate proportionally to his similarity to him.
To inject prior knowledge on who the {\itshape papabili} are, cardinals may only vote for the following candidates in the first round: Ambongo Besungu, Arborelius, Aveline, Bo, Brislin, Burke, Eijk, Erdo, Filoni, Koch, Muller, Parolin, Pizzaballa, Ranjith, Sarah, Sturla, Tagle, Tolentino de Mendonça and Zuppi. Other candidates may however emerge in the following rounds. Specifically, the probability that $i$ casts his initial vote for $j$ is
\begin{equation}
    p^{(0)}_{ij} = \frac{S_{ij}^\gamma q_j  }{\sum_{k} S_{ik}^\gamma q_k},
\end{equation}
where $\gamma\geq 1$ is a sharpening parameter to translate the similarity generated from the embedding into a probability of voting, and $q_i=1$ if $i$ is in the initial list of {\itshape papabili}, zero otherwise.

After the first round, anyone can be voted and two mechanisms contribute to voting. One translates the individual affinity (as in the first round), this time to all cardinal electors, and is encoded as follows:
\begin{equation}
    p^{(1)}_{ij} = \frac{S_{ij}   }{\sum_{k} S_{ik} }.
\end{equation}

The second mechanism encodes the fact that one is influenced by the voting choices of those similar to him. Let us assume that $Z_{a,ij}$ is $=1$ if cardinal $i$ votes for $j$ during round $a$. Then, at round $a$ the second mechanism is encoded in the following probability:
\begin{equation}
    p^{(2)}_{a+1,ij} = \frac{ \left(S\,Z_a\right)_{ij}   }{\sum_{k} \left(S\,Z_a\right)_{ik} }
\end{equation}
Then, the final probability at round $a$ is 
\begin{equation}
    p_{a,ij} = \frac{  \left[ (1-\omega)p^{(1)}_{ij} + \omega p^{(2)}_{a,ij} \right]^\gamma }{ \sum_k \left[ (1-\omega)p^{(1)}_{ij} + \omega p^{(2)}_{a,ij} \right]^\gamma}. 
\end{equation}
Here, $\gamma$ is the usual sharpening parameter and $\omega$ is a mixture parameter that tunes the probability of feeling the influence of others vs voting according direct similarity.

Conclave rounds are then simulated by sampling from these probabilities. Rounds are performed either until a candidate is elected, i.e., reaches a $2/3$ majority of votes, or they reach the cap of $512$. While the number of rounds of the upcoming conclave is not known, it should be noted that the two previous conclaves of the 21st century had 4 and 5 ballots, and no conclave in the 20th century lasted longer than for 14 ballots. Also, we do not model the runoff system after the 13th day instated by Benedict XVI's {\itshape motu proprio} of 2007 \cite{motuproprio}.

We explored the parameters in the range $\gamma\in[1.5, 4]$ and $\omega\in[0.25, 1]$ and simulated $10,000$ conclaves for each parameter configuration. We discarded the parameter configurations leading to the conclave not converging more than $1\%$ of the time, as they led to unrealistically long conclaves. This meant in practice discarding configuration with both $\gamma, \omega$ low. The other gave reasonable number of ballots, mostly between 4 and 10. The ranges shown in the figures represent the parameter variability.

We simulated conclaves on each similarity map (for each topic) and then aggregated the similarity matrices and simulated the voting process to produce the aggregate predictions.

\section{Results}
\begin{figure*}
    \centering
    \includegraphics[width=\linewidth]{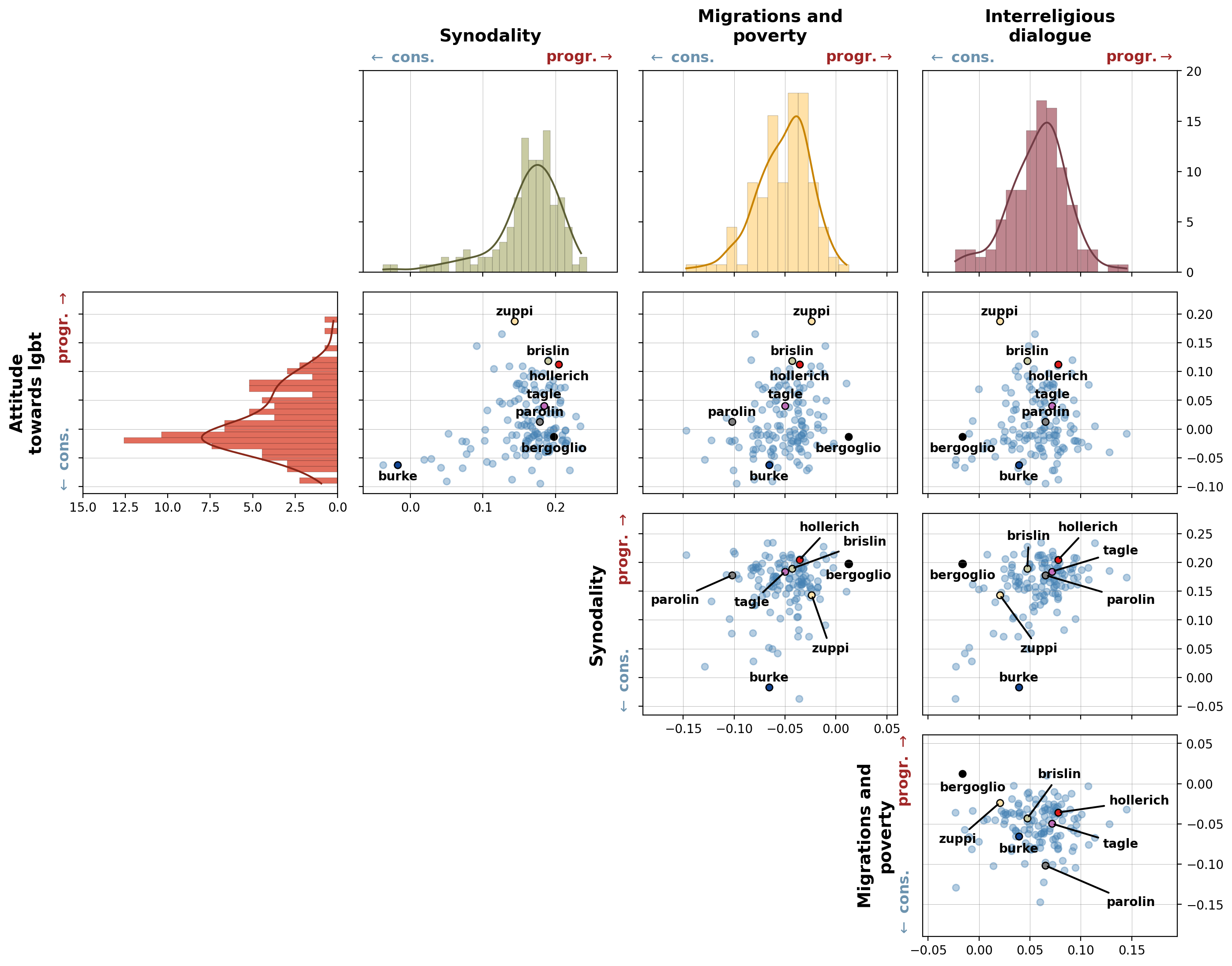}
    \caption{Distributions and bivariate scatterplots of the cardinals' similarity scores across the four selected themes (Attitude towards LGBT, Synodality, Migrants and Poverty, Interreligious Dialogue). A selected set of cardinals is labeled within the scatterplots to aid interpretation, including the late Pope Francis, whose position serves as a reference point for comparison.}
    \label{fig:scatter}
\end{figure*}

Figure~\ref{fig:scatter} displays the distribution of similarity scores for all cardinals across the four selected themes. The marginal distributions show the spread of ideological alignment for each theme individually, while the off-diagonal scatterplots illustrate pairwise relationships between themes. Overall, the College of Cardinals exhibits a mildly progressive orientation on \textit{Synodality} and \textit{Interreligious dialogue}, while tending toward slightly more conservative positions on Attitude towards LGBT and Migrants and Poverty. To aid interpretation, we have highlighted a subset of individual cardinals within the figure as illustrative reference points.

\begin{figure}[h]
    \centering
    \includegraphics[width=\linewidth]{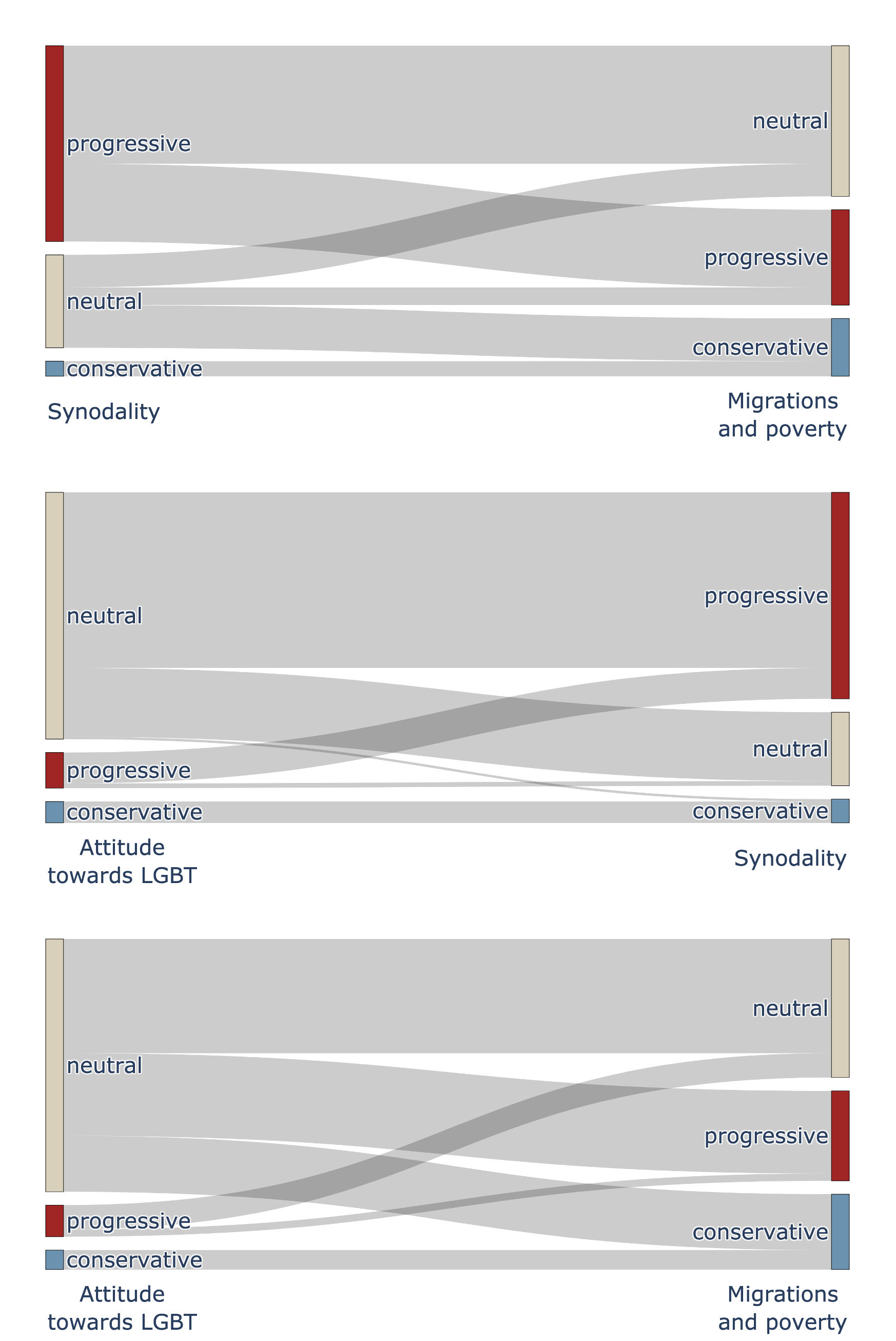}
    \caption{Sankey diagrams showing transitions between cardinal clusters across selected themes, based on spectral clustering of similarity matrices. Clusters are labeled as progressive, neutral, or conservative according to their average ideological alignment. }
    \label{fig:sankey}
\end{figure}

Figure~\ref{fig:sankey} shows the fluxes between cardinal clusters across selected themes, based on spectral clustering of the similarity matrices. Each diagram tracks transitions from ideological clusters on one theme (left) to another (right), using three categories—progressive, neutral, and conservative—defined by aggregated alignment scores.

In the top panel (Synodality → Migrants and Poverty), most cardinals progressive on Synodality remain progressive or neutral on the socioeconomic axis, indicating partial continuity. Some neutral cardinals, however, shift toward conservative positions, suggesting that moderate views on governance do not always translate into progressive stances on economic issues.

In the middle panel, the neutral cluster on LGBT dominates and disperses across all positions on Synodality, including progressive. Even cardinals initially classified as progressive or conservative on LGBT issues are often absorbed into broader neutral or progressive clusters on governance, reflecting a shared openness to synodal reform.

The bottom panel again shows a dominant neutral cluster on LGBT, spreading across all socioeconomic positions. While no strong coupling emerges between the two themes, a notable subset of progressive cardinals maintains consistent alignment across both, hinting at limited ideological cohesion within the reform-oriented bloc.

Together, the diagrams highlight both areas of consistency—especially among progressives—and significant thematic independence, underscoring the multidimensional nature of ideological positions within the College.

\begin{figure*}[htbp]
    \centering
    
    \begin{subfigure}{.85\textwidth}
        \centering
        \includegraphics[width=\textwidth]{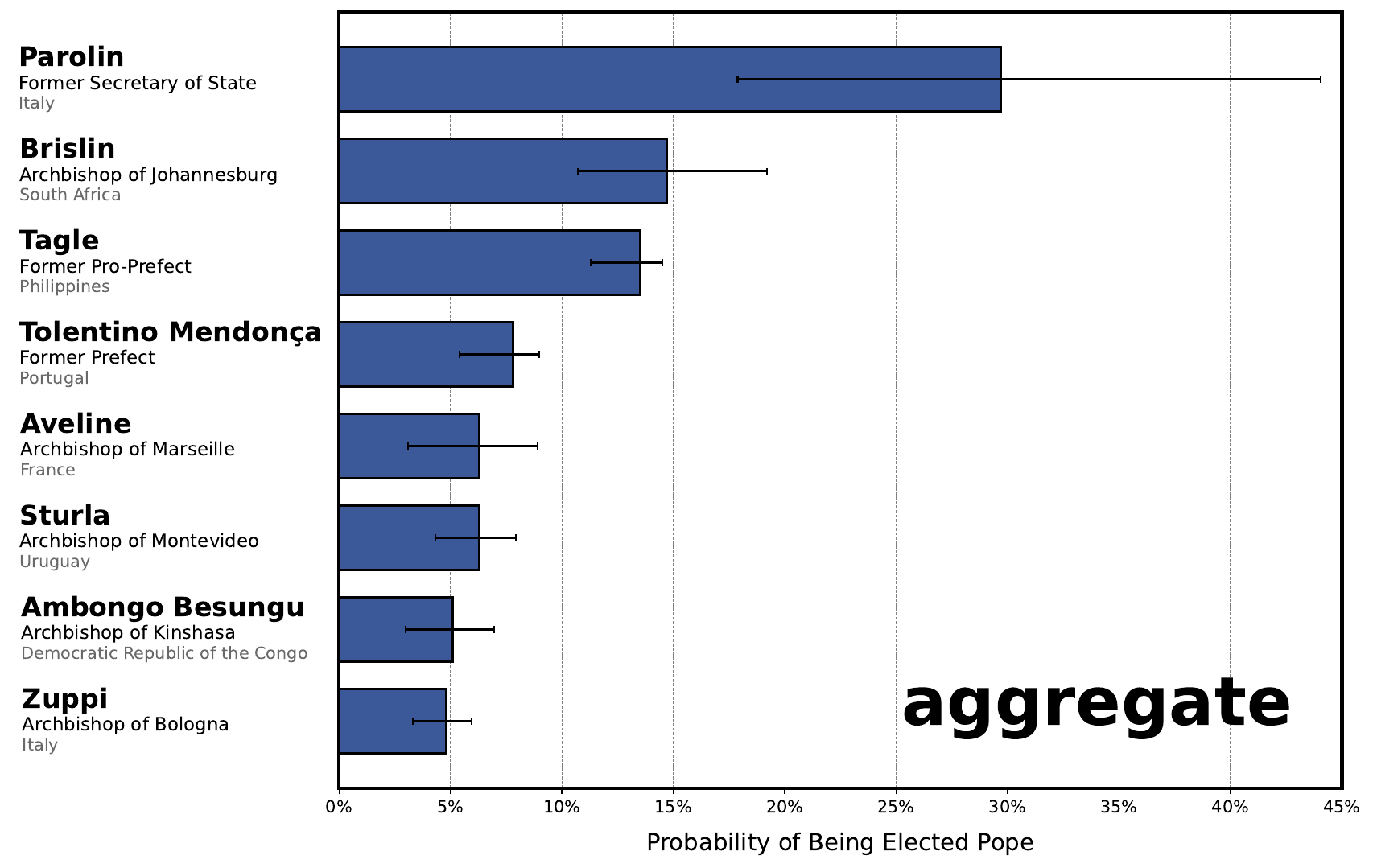}
        \label{fig:panel-a}
    \end{subfigure}
    
    \vspace{0.5em}

    \begin{subfigure}{0.49\textwidth}
        \centering
        \includegraphics[width=\linewidth]{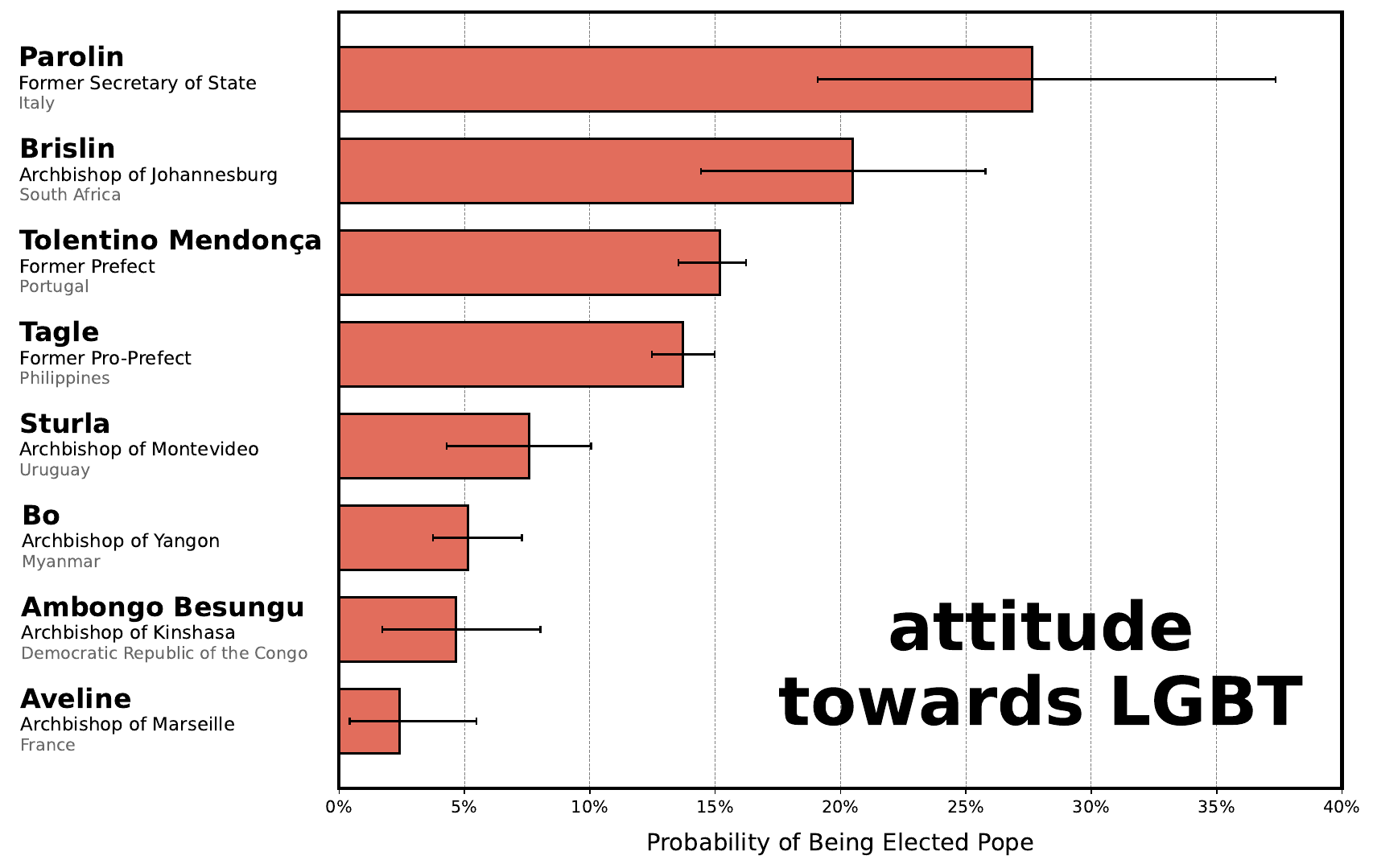}
        \label{fig:panel-b}
    \end{subfigure}
    \hfill
    \begin{subfigure}{0.49\textwidth}
        \centering
        \includegraphics[width=\linewidth]{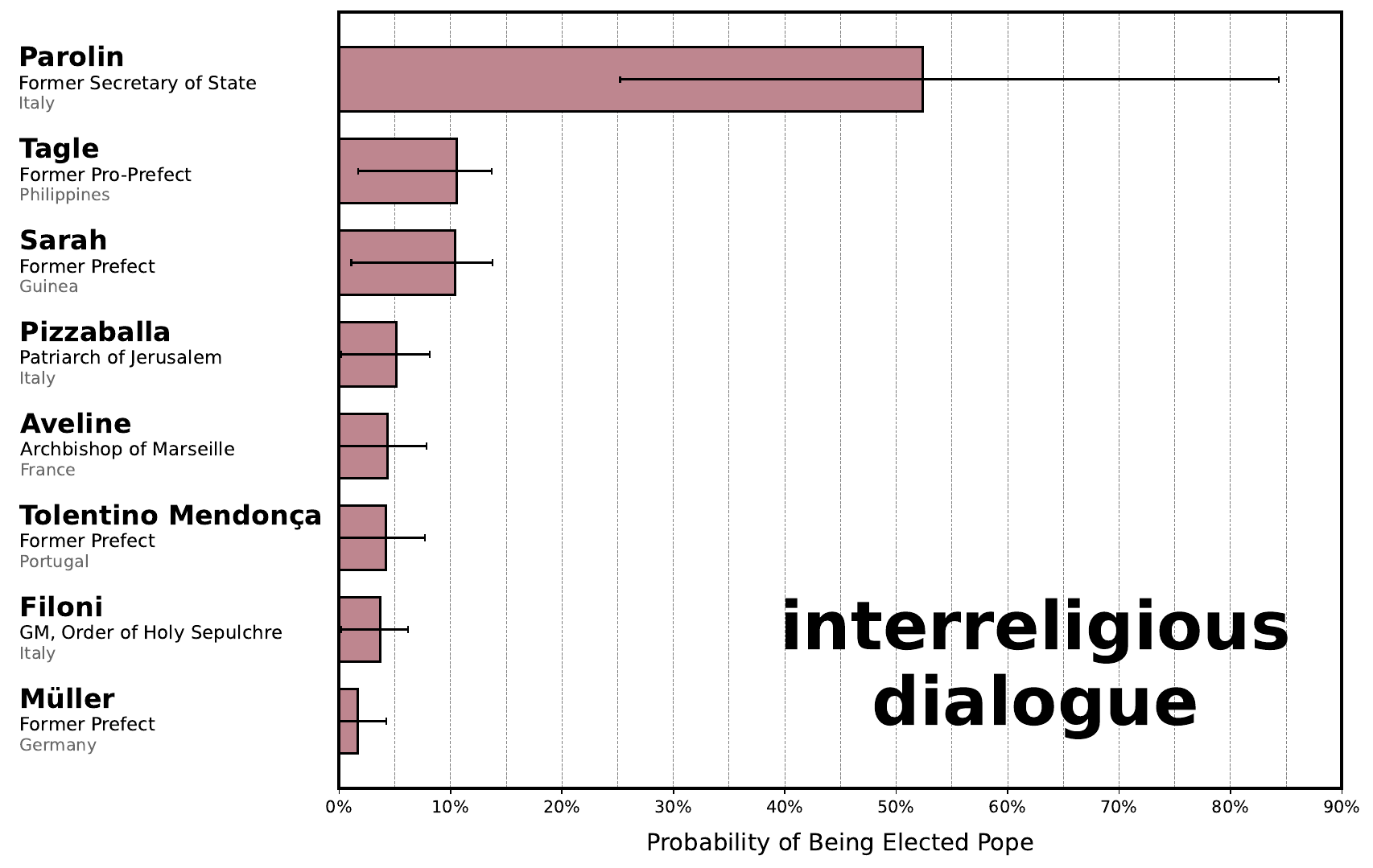}
        \label{fig:panel-c}
    \end{subfigure}

    \vspace{0.5em}

    \begin{subfigure}{0.49\textwidth}
        \centering
        \includegraphics[width=\linewidth]{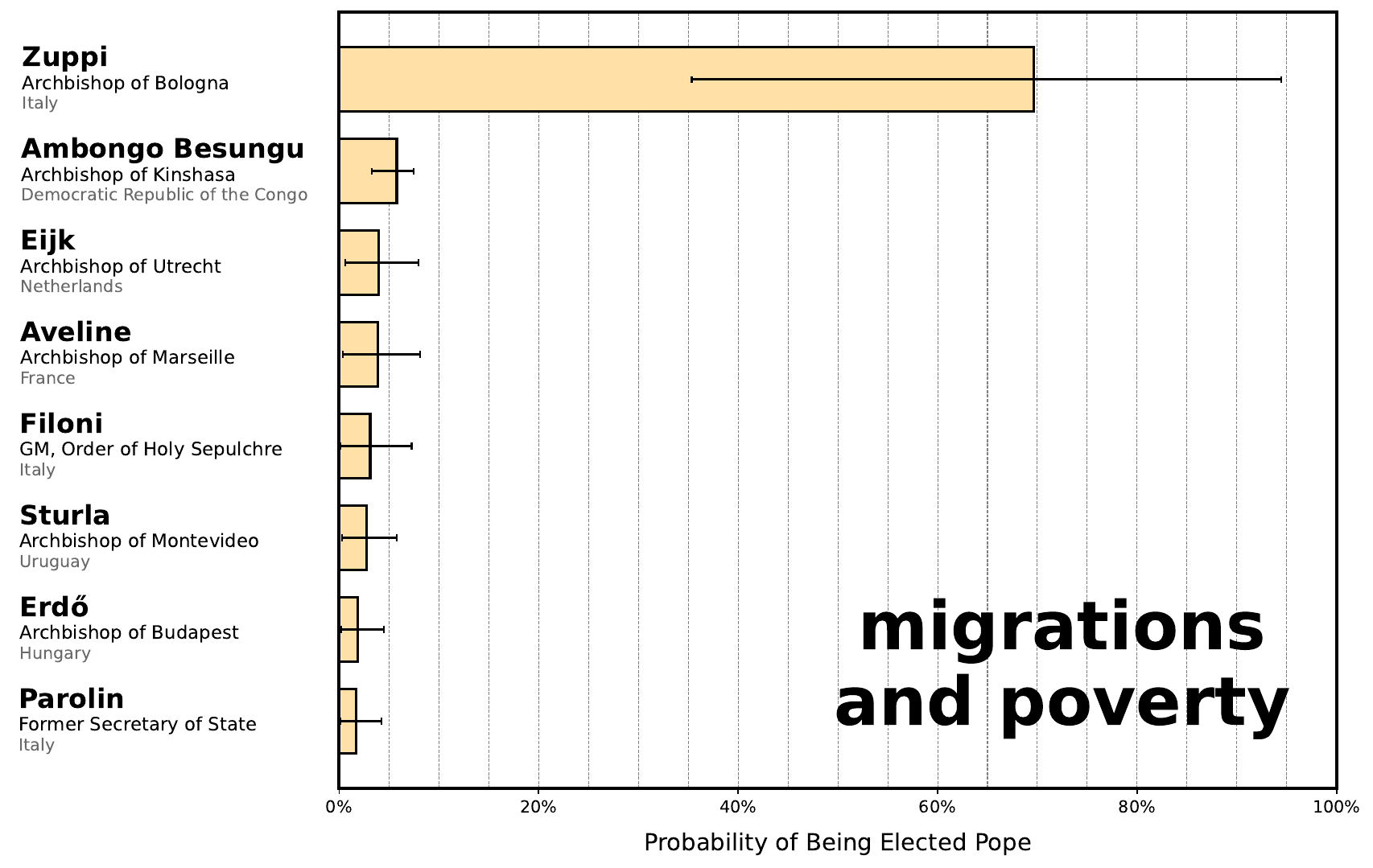}
        \label{fig:panel-d}
    \end{subfigure}
    \hfill
    \begin{subfigure}{0.49\textwidth}
        \centering
        \includegraphics[width=\linewidth]{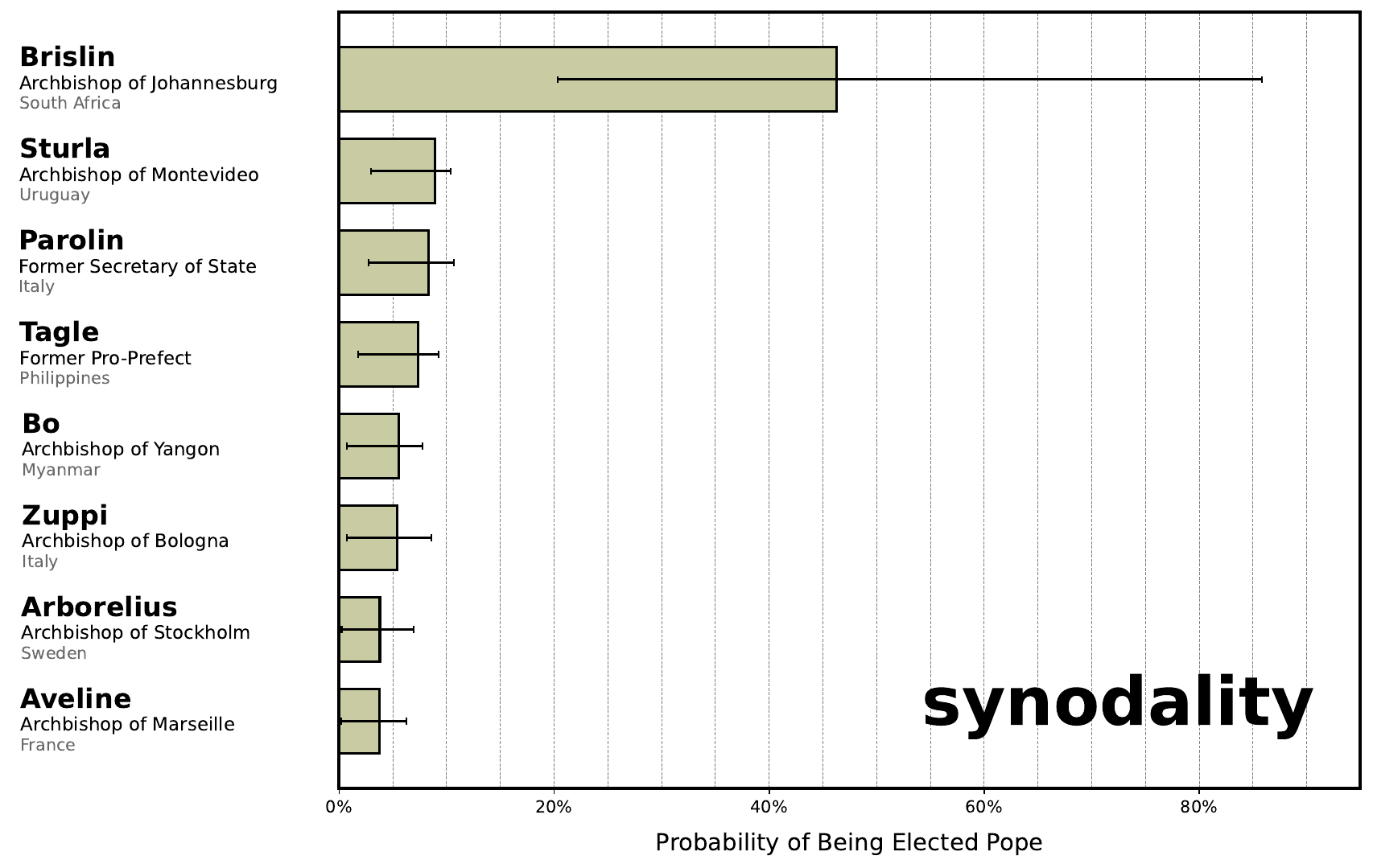}
        \label{fig:panel-e}
    \end{subfigure}

    \caption{Probability of election to the papacy for each cardinal under different topic weighting scenarios. Each conclave simulation was run 10,000 times for each set of parameter values. The top panel shows the scenario in which all topics are equally weighted. In the bottom panels, each topic is assigned maximum weight in turn, with all others set to zero. Error bars indicate the minimum and maximum probabilities observed across the parameter samples. Only the top eight cardinals by election probability are shown for each scenario.}
    \label{fig:composite}
\end{figure*}

Finally, the results of the voting simulations are reported in Figure~\ref{fig:composite}. Here, the bars present the estimated probabilities of being elected Pope for all cardinals, ranked from highest to lowest. Probabilities are reported both by theme and in aggregate form. The top panel shows the aggregate ranking, where all themes are considered jointly; under this model, Cardinal Parolin is the leading candidate and would be elected, followed by Brislin, Tagle, and Tolentino Mendonça. The four lower panels show conditional probabilities assuming that one specific theme is regarded as the most pressing issue in the Conclave. That is, each panel reflects the hypothetical outcome if the election were primarily guided by alignment on that theme.

Under \textit{Attitude towards LGBT}, Parolin again emerges as the leading candidate, closely followed by Brislin and Tolentino Mendonça, suggesting that moderate-progressive figures are favored when this theme is prioritized. When \textit{Interreligious Dialogue} is considered decisive, card. Parolin becomes the overwhelming frontrunner, indicating broad alignment with his stance in that domain. The \textit{Migrations and Poverty} panel produces a different result: card. Zuppi dominates, followed by Ambongo Besungu—both cardinals known for strong public engagement on socio-economic issues. Finally, under Synodality, Brislin ranks first, ahead of Sturla and Parolin, showing that his profile resonates most with governance-focused reform. These variations underscore the sensitivity of the election outcome to the thematic focus of the Conclave, though a few figures —such as card. Parolin— maintain high viability across multiple scenarios.

\section{Conclusion}
In this study, we explored the internal diversity of the College of Cardinals by applying cross-encoding techniques to analyze both pairwise similarity between cardinals and their alignment with predefined ideological stances. By focusing on four central themes—\textit{Attitude towards LGBT, Synodality, Migrants and Poverty, and Interreligious Dialogue}—we constructed semantic similarity matrices that reveal how closely each cardinal’s discourse aligns with others on a given topic. Spectral clustering on these matrices uncovered distinct thematic groupings, with some cardinals consistently forming progressive-leaning clusters, while others exhibited more variable or theme-specific patterns. 
In parallel, we quantified each cardinal’s position along a progressive–conservative axis by computing alignment scores relative to synthetic reference texts representing both ends of the spectrum for each theme. This method provided a continuous measure of ideological orientation and allowed for comparisons not only between individuals but also across themes. Some cardinals—such as Parolin and Brislin—maintained relatively stable moderate-to-progressive profiles across most issues, while others—such as Zuppi or Sarah—stood out as distinctly aligned with specific themes but not across the board.
Notably, \textit{Synodality} and \textit{Interreligious Dialogue }showed broader consensus and a stronger progressive core, whereas \textit{Attitude towards LGBT} and \textit{Migrations and Poverty} revealed more ideological dispersion, with sizable neutral clusters and sharper polarizations.

Finally, we translated these ideological profiles into predicted voting outcomes using a simple probabilistic model that assumes thematic alignment plays a decisive role in determining papal viability. We simulated elections under the assumption that each theme, in turn, is regarded as the dominant concern of the Conclave. The resulting theme-specific rankings highlight how different cardinals rise to prominence depending on the prevailing issue. Under \textit{Migrtions and Poverty}, for instance, card. Zuppi becomes the frontrunner, while \textit{Synodality} favors card. Brislin. When all themes are considered jointly, card. Parolin emerges as the overall leading candidate, combining broad-based alignment with centrist appeal. He is followed by others mixing high-ranking and outsider names such as cards. Tagle, Brislin, and Tolentino Mendonça.

Needless to say, this report has no ambition of becoming a scientific study. To us it has been a fun and stimulating exercise and we share hoping that fellow nerds will find it interesting. Notwithstanding, do not hesitate to reach out to us if you have comments and suggestions on the methodology.

\section{\textit{Addendum}: Review of the Model and Considerations Following the Election}

We briefly report a post-election comment on the performance of our model. 

\begin{figure*}
    \centering
    \includegraphics[width=\linewidth]{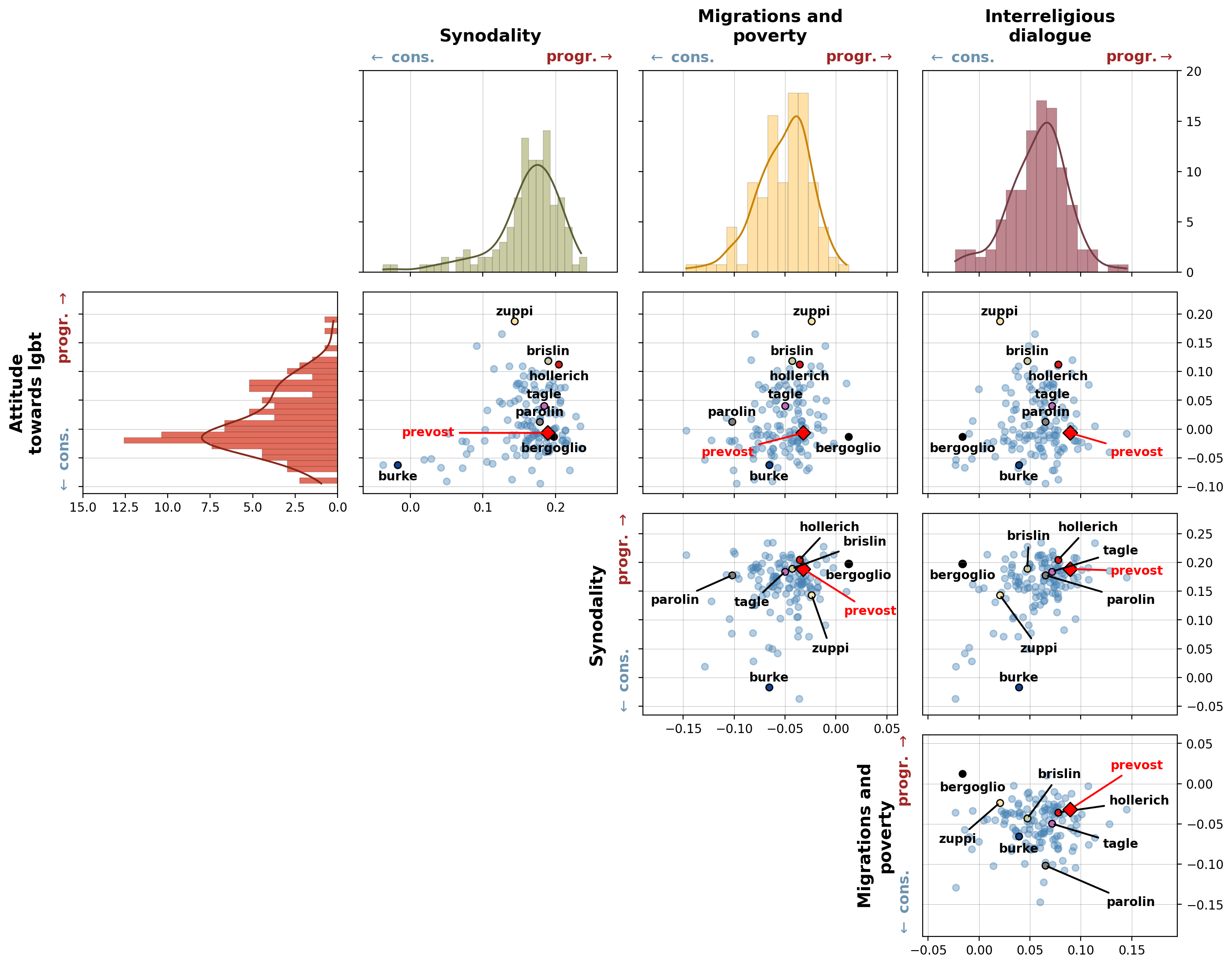}
    \caption{Distributions and bivariate scatterplots of the cardinals as in Figure~\ref{fig:scatter}, including Prevost in the prior.}
    \label{fig:scatter1}
\end{figure*}

Clearly, our predictions missed Robert F. Prevost, who was elected pope as Leo XIV. Further analysis showed that the multilayer mapping effectively captured cardinal ideological affinity, stratifying by the different challenges facing the Catholic Church. Interestingly, as shown in Figure~\ref{fig:scatter1}, our ideological mapping places the newly elected pope as a centrist in all topics except \textit{Interreligious Dialogue}, where he is moderately progressive.

The voting model, limited by the absence of a validation set, inevitably depended on prior knowledge, most notably, the initial list of \textit{papabili} used in the first round of voting. In our forecast, that list did not include cardinal Prevost, which substantially reduced his estimated likelihood of election. This dependence on untested assumptions is an inherent limitation in the absence of empirical data for calibration.

In addition, the voting model relies on ideological similarity across four selected themes and does not include any geographical or geopolitical information. Attempts to treat geographical affinity in the same way as the other themes yielded mostly noise and negligible signal, and was therefore excluded from the final model. However, this omission represents a significant shortcoming, as spatial and geopolitical factors are likely to play an important role in conclave dynamics. Incorporating them appropriately will be necessary to obtain more reliable results.

\begin{figure*}[htbp]
    \centering
    
    \begin{subfigure}{.85\textwidth}
        \centering
        \includegraphics[width=\textwidth]{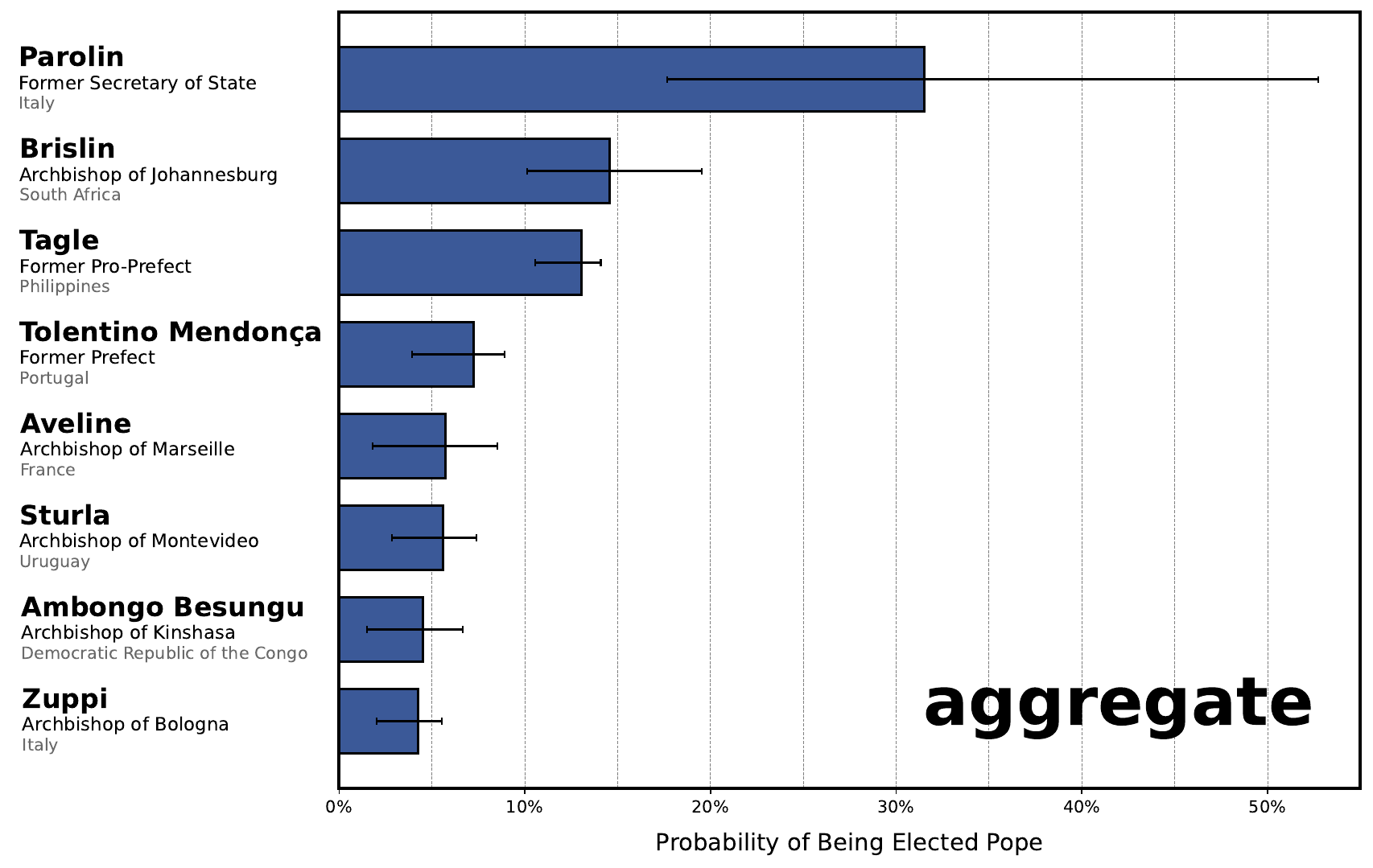}
        \label{fig:panel-a}
    \end{subfigure}
    
    \vspace{0.5em}

    \begin{subfigure}{0.49\textwidth}
        \centering
        \includegraphics[width=\linewidth]{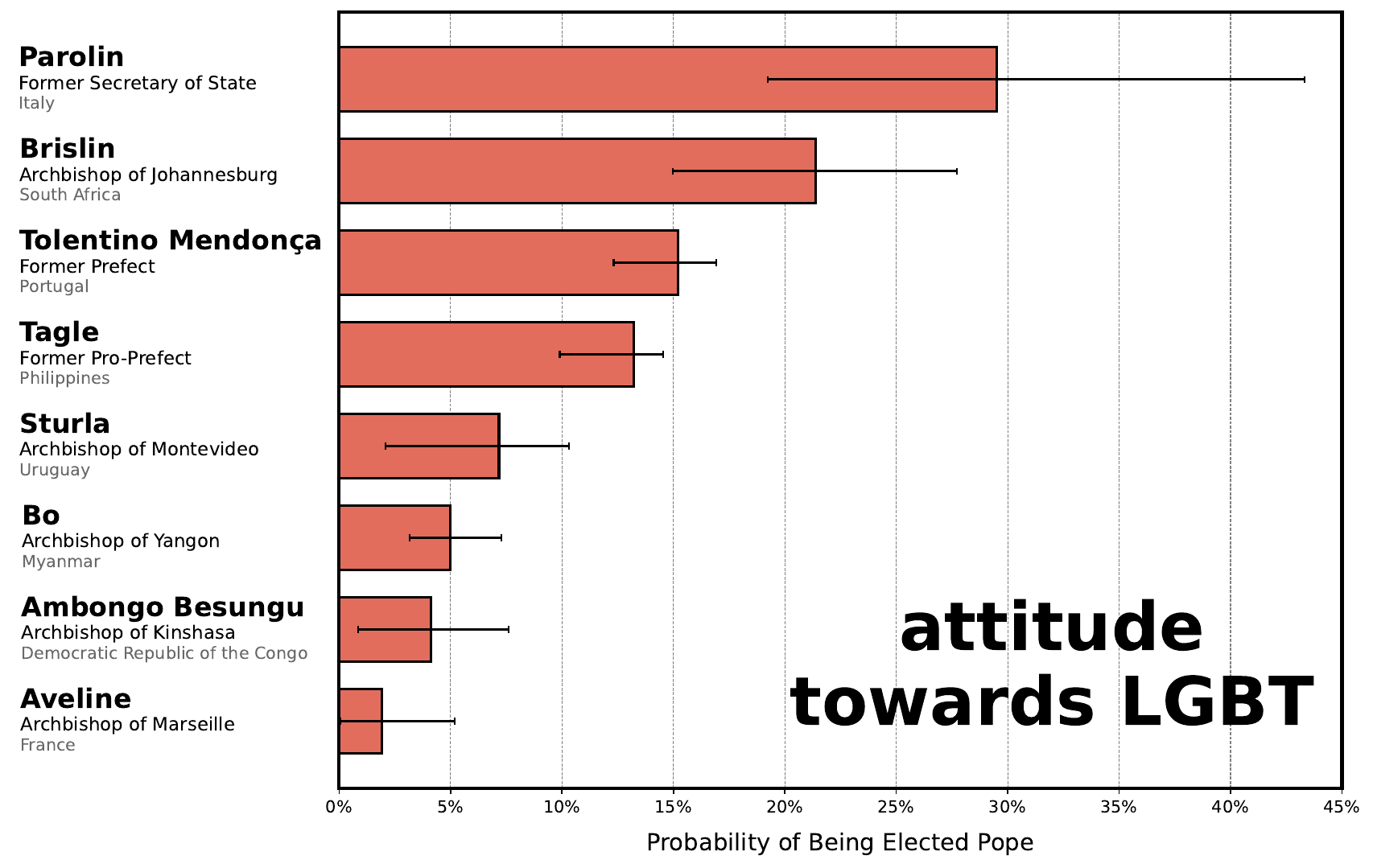}
        \label{fig:panel-b}
    \end{subfigure}
    \hfill
    \begin{subfigure}{0.49\textwidth}
        \centering
        \includegraphics[width=\linewidth]{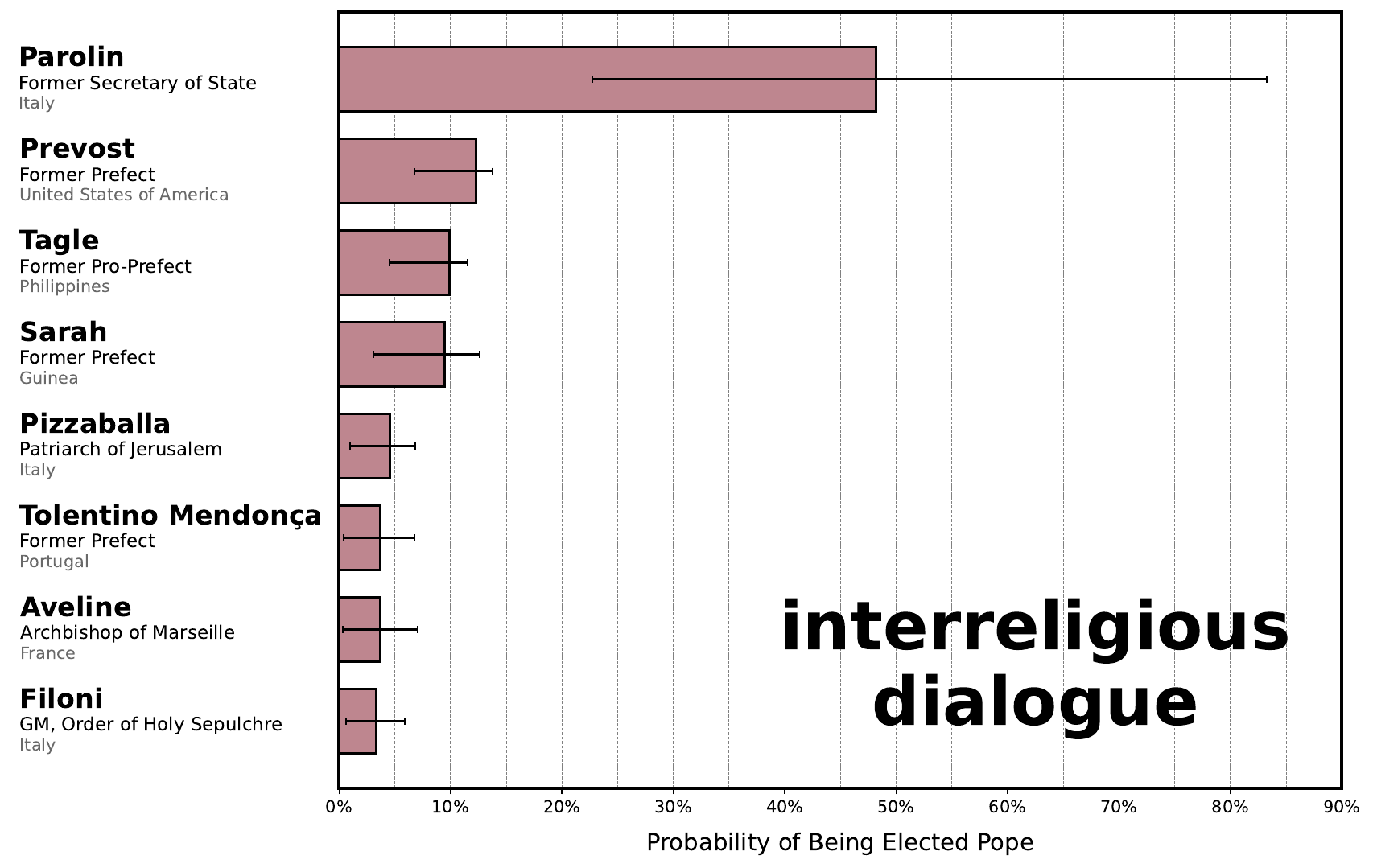}
        \label{fig:panel-c}
    \end{subfigure}

    \vspace{0.5em}

    \begin{subfigure}{0.49\textwidth}
        \centering
        \includegraphics[width=\linewidth]{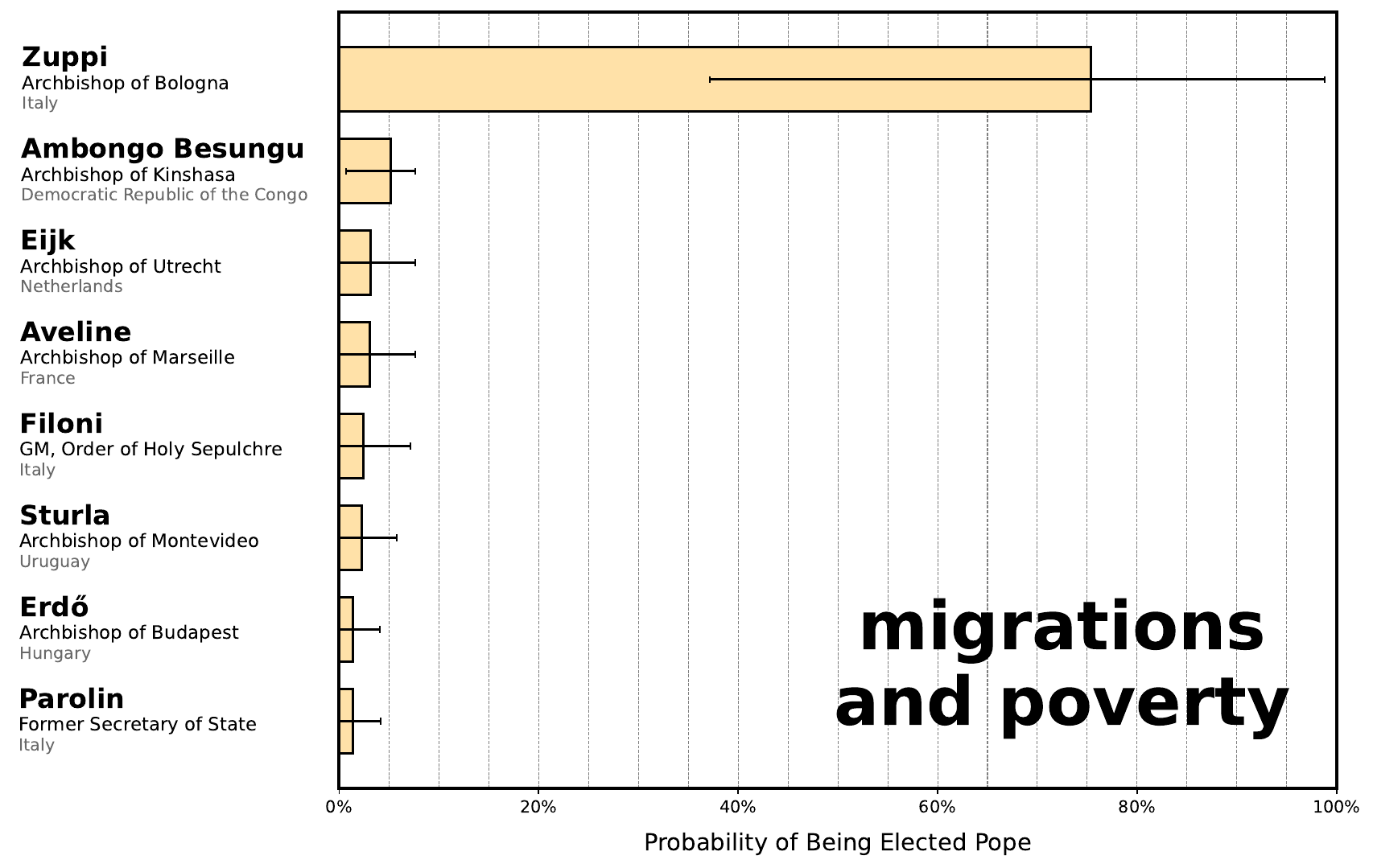}
        \label{fig:panel-d}
    \end{subfigure}
    \hfill
    \begin{subfigure}{0.49\textwidth}
        \centering
        \includegraphics[width=\linewidth]{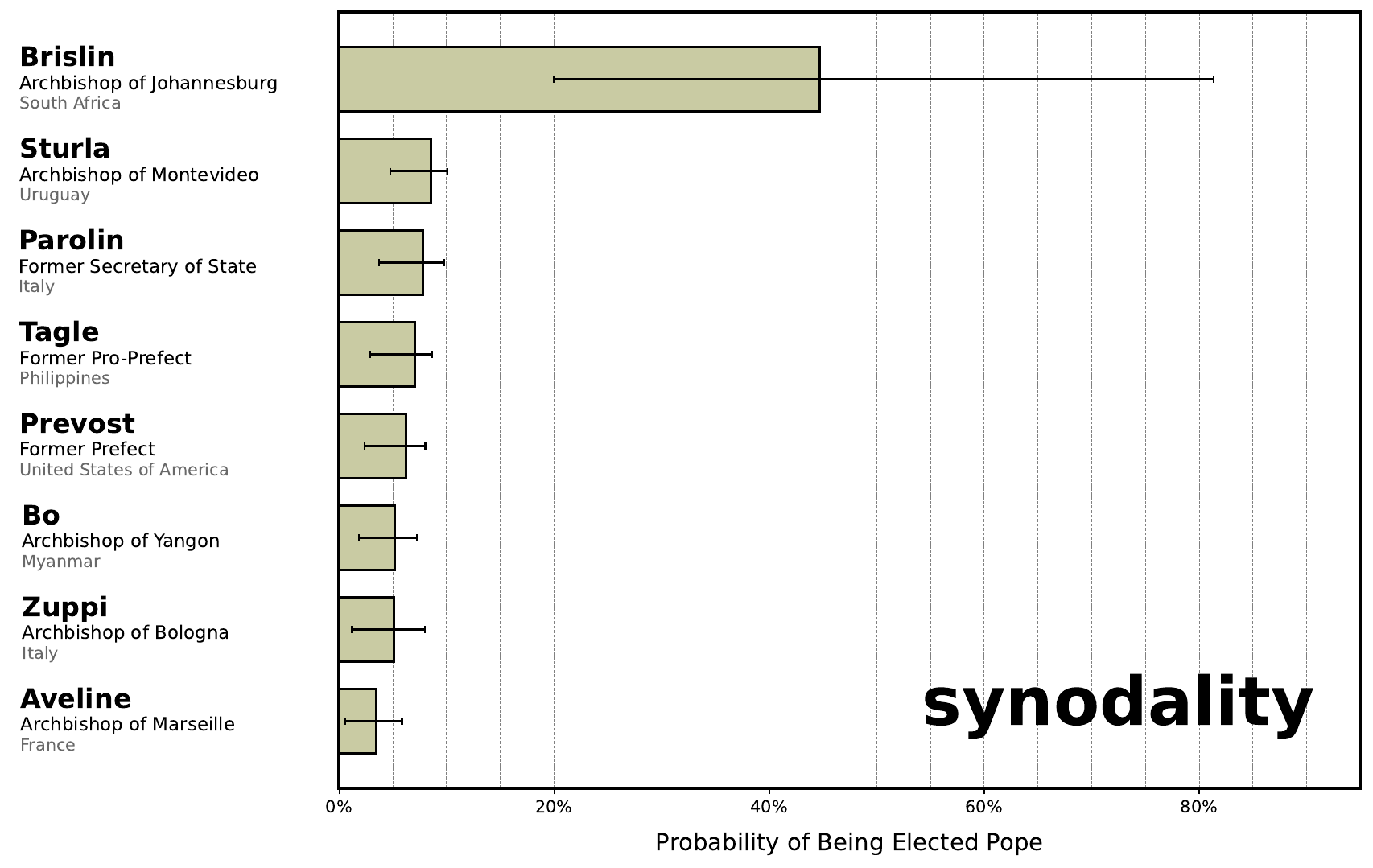}
        \label{fig:panel-e}
    \end{subfigure}

    \caption{Probability of election to the papacy as in Figure~\ref{fig:composite}, including Prevost in the prior.}
    \label{fig:composite1}
\end{figure*}

We re-ran the voting model including Prevost in the prior list. While this still gave him no substantial chance of being elected in the aggregate scenario, he emerged as a close second in the theme of dialogue, as shown in Figure~\ref{fig:composite1}, a theme that featured prominently in his first speech.

\end{document}